\documentclass[reprint,aps,amsmath,amssymb,amsfonts,superscriptaddress]{revtex4-2}

\usepackage{braket}
\usepackage{color}
\usepackage[normalem]{ulem}
\usepackage{bm}

\def\beq{\begin{equation}}
\def\eeq{\end{equation}}
\def\bea{\begin{eqnarray}}
\def\eea{\end{eqnarray}}
\def\nn{\nonumber}
\def\nl{\nonumber\\}
\def\bd{B_d^0}

\def\pewcp{P_{EW}^{\prime C}}
\def\pewp{P'_{EW}}
\def\btopik{B \to \pi K}
\def\ft{\mathfrak f}
\def\bsll{b \to s \ell^+ \ell^-}

\begin{document}

\title{\boldmath  Axion-like particles resolve the $\btopik$ and $g-2$ anomalies}

\author{Bhubanjyoti Bhattacharya}
\affiliation{Department of Natural Sciences, Lawrence Technological University, Southfield, MI 48075, USA}
\email{bbhattach@ltu.edu}

\author{Alakabha Datta}
\affiliation{Department of Physics and Astronomy, \\
108 Lewis Hall, University of Mississippi, Oxford, MS 38677-1848, USA}
\email{datta@phy.olemiss.edu}

\author{Danny Marfatia}
\affiliation{Department of Physics and Astronomy, \\
2505 Correa Rd, University of Hawaii at Manoa, Honolulu, HI 96822, USA}
\email{dmarf8@hawaii.edu}

\author{Soumitra Nandi}
\affiliation{Department of Physics, Indian Institute of Technology, Guwahati 781 039, India}
\email{soumitra.nandi@iitg.ac.in}

\author{John Waite}
\affiliation{Department of Physics and Astronomy, \\
108 Lewis Hall, University of Mississippi, Oxford, MS 38677-1848, USA}
\email{jvwaite@go.olemiss.edu}

\begin{abstract}
We offer a new solution to an old puzzle in the penguin-dominated $\btopik$ decays. The puzzle is the
inconsistency among the measurements of the branching ratios and CP asymmetries of the four $\btopik$ decays: $B^+ \to \pi^+ K^0$, $B^+\to \pi^0 K^+$, $\bd\to\pi^- K^+$, $\bd \to \pi^0 K^0$. We solve the $\btopik$
puzzle by considering the effect of an axion-like particle (ALP) that mixes with the $\pi^0$ and
has mass close to the $\pi^0$ mass.  We show that the ALP can also explain the anomalies in the electron and muon anomalous magnetic moments.
\end{abstract}


\preprint{
{\flushright
UMISS-HEP-2021-02 \\
}}

\maketitle

\textbf{\emph{Introduction}.--} The $b$-quark system is known to be an excellent place to test the Standard Model
(SM) as well as models of New Physics (NP). Flavor Changing Neutral Current (FCNC) processes like $b\to s$ penguin
decays are ideal places to look for NP. The FCNC decays in the $B$ system have been studied in detail by experiments
over the years. In recent times, measurements in the FCNC semileptonic $\bsll$ decays have revealed discrepancies
with SM predictions. These discrepancies, or anomalies, have been widely studied over most of the last decade.
Almost a decade before the semileptonic $\bsll$ anomalies arose, another anomaly in non-leptonic $B$-meson decays
dominated by $b\to s$ penguins had attracted a great deal of interest. The anomaly was in the CP violation measurement
of $\btopik$ decays where an inconsistency was observed and this was called the ``$\btopik$ puzzle'' \cite{Buras:2003yc,Buras:2003dj,Buras:2004ub}. The amplitudes of the four $\btopik$ decays, $B^+\to\pi^+ K^0$
(designated as $+0$ below), $B^+ \to \pi^0 K^+$ ($0+$), $\bd \to \pi^-K^+$ ($-+$), and $\bd \to \pi^0 K^0$ ($00$),
are related by a single isospin relationship,
\beq
\sqrt{2} A^{00} + A^{-+} = \sqrt{2} A^{0+} + A^{+0}\,. \label{eq:isorel}
\eeq
In these decays, experiments measure nine observables: the four branching ratios, the four direct CP asymmetries $A_{CP}$,
and the mixing-induced indirect CP asymmetry $S_{CP}$ in $\bd\to\pi^0K^0$. Expressing the $\btopik$ decays in terms of
topological amplitudes one can perform a fit to obtain the SM as well as the NP amplitudes \cite{Baek:2004rp}. As new
experimental numbers were reported, updated fits were performed in Refs.~\cite{Baek:2007yy, Baek:2009pa, Beaudry:2017gtw, Fleischer:2017vrb, Fleischer:2018bld}. Although the fits revealed a strong hint of NP in these decays, complicated strong
dynamics made it difficult to draw a definite conclusion.

In this paper we explore the possibility that a light pseudoscalar particle close to the pion mass can solve the $\btopik$ puzzle.
The key observation is that in the $\btopik$ set of decays, the discrepancies from the SM predictions involve modes with a $\pi^0$
in the final state. The basic idea to solve the $\btopik$ puzzle is to assume that there is a light pseudoscalar particle, $a$,
that mixes with the $\pi^0$. In our model, an FCNC $B\to Ka$ amplitude is generated through the usual top-penguin diagram followed
by the $a$ mixing with the $\pi^0$ to produce a new contribution to the $B \to K \pi^0$ amplitudes. We then show that this new amplitude can solve the $\btopik$ puzzle while being consistent with constraints from various other processes. We point out that the ALP can also solve the $(g-2)_{\mu,e}$ anomalies via its couplings to leptons and photons.

\textbf{\emph{\bm{$B\to\pi K$} puzzle}.--} We begin by explaining the $\btopik$ puzzle by following the discussion in Ref. \cite{Beaudry:2017gtw}. Within the diagrammatic approach \cite{Gronau:1994rj, Gronau:1995hn}, $B$-decay amplitudes are expressed in terms of six diagrams. The $\btopik$ decay amplitudes are
\bea
\label{piKamp}
A^{+0} &=& -P'_{tc} + P'_{uc} e^{i\gamma} -\frac13
\pewcp\,, \\
\sqrt{2} A^{0+} &=& -T' e^{i\gamma} -C' e^{i\gamma}
+P'_{tc} -~P'_{uc} e^{i\gamma} \nl
&&-~\pewp -\frac23 \pewcp\,, \\
A^{-+} &=& -T' e^{i\gamma} + P'_{tc} -P'_{uc}
e^{i\gamma} -\frac23 \pewcp\,, \\
\sqrt{2} A^{00} &=& -C' e^{i\gamma} - P'_{tc} + P'_{uc} e^{i\gamma} \nl
&&-~\pewp -\frac13 \pewcp\,.
\eea
The various diagrams are discussed in Ref. \cite{Baek:2004rp} and in the topological amplitudes above, we explicitly show the weak-phase dependence. In these decays the electroweak penguin amplitudes play an important role and it has been shown \cite{Neubert:1998pt,Neubert:1998jq,Gronau:1998fn} that, to a good approximation, the electroweak
penguins $\pewp$ and $\pewcp$ can be related to the tree-level diagrams $T'$ and $C'$ within the
SM using flavor-SU(3) symmetry:
\bea
\label{EWPrels}
\pewp & \!\!=\!\! & {3\over 4} {c_9 + c_{10} \over c_1 + c_2} R (T' +
C') \!+\!  {3\over 4} {c_9 - c_{10} \over c_1 - c_2} R (T' - C')\,, \nn\\
\pewcp & \!\!=\!\! & {3\over 4} {c_9 + c_{10} \over c_1 + c_2} R (T' +
C') \!-\!  {3\over 4} {c_9 - c_{10} \over c_1 - c_2} R (T' - C')\,,
\eea
where the $c_i$ are Wilson coefficients (WC) \cite{Buchalla:1995vs} and $R \equiv \left\vert (V_{tb}^*V_{ts}) /(V_{ub}^* V_{us}) \right\vert = 45.8$ \cite{pdg}. Following Eq.~(\ref{EWPrels}), $\pewp$ receives a relatively large contribution from $T'$ but a much smaller contribution from $C'$. In contrast, $\pewcp$ receives a relatively large contribution from $C'$ and a much smaller $T'$ contribution. In this sense, $\pewp$ and $T'$ are of roughly similar size, and so are $\pewcp$ and $C'$.

\textbf{\emph{\bm{$\btopik$} puzzle simplified}.--} Keeping the leading-order diagrams in Eq.~(\ref{piKamp}), the $\btopik$
amplitudes become
\bea
\label{reducedamps}
A^{+0} &=& -P'_{tc}\,, \nn\\
\sqrt{2} A^{0+} &=& -T' e^{i\gamma} + P'_{tc} - \pewp\,, \nn\\
A^{-+} &=& -T' e^{i\gamma} + P'_{tc}\,, \nn\\
\sqrt{2} A^{00} &=& - P'_{tc} - \pewp\,.
\eea
Consider, now, the direct CP asymmetries of $B^+\to\pi^0 K^+$ and $\bd\to\pi^- K^+$. A direct CP asymmetry
is generated by the interference of two amplitudes with nonzero relative weak and strong phases. In $A^{-+}$, $T'$-$P'_{tc}$ interference leads to a direct CP asymmetry. Note that $\pewp$ and $P'_{tc}$ have the same weak phase ($=0$). As discussed earlier, $\pewp \propto T'$ once we neglect $C'$ (see Eq.~\ref{EWPrels}). Therefore, if we assume that $\pewp$ and $T'$ have a similar strong phase, the contribution to direct CP-asymmetry in $A^{0+}$ can be assumed to be originated from the interference of $T'$-$P'_{tc}$. This means, to leading order in $|T'|/|P'_{tc}|$, we expect $A_{CP}(B^+ \to \pi^0 K^+) = A_{CP}(\bd \to \pi^- K^+)$.

The latest $\btopik$ measurements are shown in Table \ref{tab:data}. Not only are $A_{CP}(B^+ \to \pi^0 K^+)$
and $A_{CP}(\bd \to \pi^-K^+)$ not equal, they are of opposite sign!  Experimentally, $(\Delta A_{CP})_{\rm exp}
= A_{CP}^{0+} - A_{CP}^{-+} = (10.8 \pm 1.6) \%$ which differs from 0 by $6.5\sigma$. We have performed a fit to data with the SM parameters. The fit is of poor quality, as we show below. This is a simplified version of the $\btopik$ puzzle.
\begin{table}[t]
\center
\begin{tabular}{|c|c|c|c|}
\hline
Decay&BR($\times10^{-6}$)~\cite{HFLAV:2016hnz}&$A_{CP}$&$S_{CP}$~\cite{HFLAV:2016hnz}\\
\hline
$B^+\rightarrow\pi^+K^0$&$23.79\pm0.75$&$-0.017\pm0.016$~\cite{pdg}&\\
$B^+\rightarrow\pi^0K^+$&$12.94\pm0.52$&$0.025\pm0.016$~\cite{LHCb:2020dpr}&\\
$B^0_d\rightarrow\pi^-K^+$&$19.57\pm0.53$&$-0.084\pm0.004$~\cite{LHCb:2018pff}&\\
$B^0_d\rightarrow\pi^0K^0$&$9.93\pm0.49$&$-0.01\pm0.10$~\cite{HFLAV:2016hnz}&$0.57\pm0.17$\\
\hline
\end{tabular}
\caption{CP-averaged branching ratios, direct CP asymmetries $A_{CP}\equiv[{\rm{BR}}(\bar{B} \to \bar{F})-{\rm{BR}}(B \to F)]/[{\rm{BR}}(\bar{B} \to \bar{F})+{\rm{BR}}(B \to F)]$ (with final states $F\,, \bar{F}$), and
  mixing-induced CP asymmetry $S_{CP}$ (if applicable) for the four
  $\btopik$ decay modes.}
  %\end{centering}
\label{tab:data}
\end{table}

\textbf{\emph{ALPs}.--} Axions and axion-like particles have been extensively studied since the introduction of the axion to solve the strong-CP problem \cite{Peccei:1977hh, Peccei:1977ur, Weinberg:1977ma,Wilczek:1977pj}. For our purpose, we assume that there is a pseudoscalar ALP $a$, that is a pseudo-Nambu-Goldstone boson, emerging from the breaking of some global $U(1)$ symmetry. We write the flavor-conserving Lagrangian for $a$ at low energy as
\bea
\mathcal{L}_{a} &=& \frac{1}{2}(\partial_\mu a)^2 - \frac{1}{2}m_{a}^2 a^2 - i \sum_{f=d,l} \xi_f \frac{m_f}{\ft} \bar{f}\gamma_5 f  a \nl
&& -~i \! \sum_{f=u} \!  \eta_f \frac{m_f}{\ft}\bar{f}\gamma_5f a  - \frac{1}{4} \kappa a F_{\mu \nu} {\widetilde{F}}^{\mu \nu}\,,
\label{L_a}
\eea
where $\ft$ is the ALP decay constant, and the dual electromagnetic field tensor is  ${\widetilde{F}}^{\mu \nu}={1\over2} \epsilon^{\mu\nu\alpha\beta}F_{\alpha\beta}$. The last term of Eq.~(\ref{L_a}) reproduces the anomalous $\pi^0 \gamma\gamma$ coupling if $a$ and $\kappa$ are replaced by $\pi^0$ and
$g_{\pi \gamma \gamma} =  \frac{\sqrt{2} \alpha}{\pi f_\pi} \sim 2.5 \times 10^{-2}$~GeV$^{-1}$ (with the neutral pion decay constant $f_\pi=130$~MeV), respectively.

We assume the ALP has properties that are desirable to solve the $\btopik$ puzzle. We take $a$ to have a mass close to the $\pi^0$ mass and require it to promptly decay to the $\gamma\gamma$ final state via its mixing with the $\pi^0$. The decay $a\to\gamma\gamma$ can occur through a direct coupling to photons or through mixing with a $\pi^0$ and so its effective
coupling is
\bea
g_{a \gamma \gamma}  = \kappa + \sin \theta  g_{\pi \gamma \gamma}\,,
\eea
where
$\sin \theta$ is the $a-\pi^0$ mixing angle. The ALP width is then $\Gamma_a = g_{a\gamma\gamma}^2m_a^3/(64\pi)$, which reduces to the $\pi^0$ width for $\kappa=0$ and $\sin\theta=1$. Assuming $\kappa \ll  \sin \theta  g_{\pi \gamma \gamma}$,
$\Gamma_a \sim  \sin^2 \theta   \Gamma_{\pi^0 }$. Since we will be interested in $ \sin{\theta} \sim 0.1$ for $m_a \sim m_{\pi^0}$~\cite{Altmannshofer:2019yji}, we have $\Gamma_a \sim 10^{-2} \Gamma_{\pi^0} \ll m_a$. Constraints on the $a \gamma \gamma$ coupling with the ALP mass near the $\pi^0$ mass have been obtained from collider and astrophysical observations~\cite{Abbiendi:2002je,Cadamuro:2011fd, Millea:2015qra, Depta:2020wmr, Ishida:2020oxl}. Our choice for the $ a \gamma \gamma$  coupling is consistent with existing constraints.  Note that our $g_{a\gamma\gamma}$ is unrelated to a tree-level $Z a \gamma$ coupling that arises in models in which the ALP couples to the $W$ and $Z$ bosons.
We assume that the ALP coupling to  photons is entirely generated by the coupling to gluons (which has  been absorbed by chiral rotation), and the coupling  to fermions.
 One-loop contributions to the $Z a \gamma$ coupling from charged fermion loops are at least three orders of magnitude smaller than the fermion couplings~\cite{Bauer:2017ris}. For $\eta_{u} , \xi_d \sim 0.01$ and $\xi_{e,\mu} \sim 0.1$, our model is unconstrained by the $Z\to\pi^0\gamma$ branching fraction~\cite{Bauer:2017ris,Jaeckel:2015jla}.

In our model, the ALP contributes to $\btopik$ decays through the $b \to s$ penguin which arises from the usual penguin loop and is divergent. We write a renormalization-group equation for the WC of the FCNC operator \cite{Chala:2020wvs, Bauer:2020jbp, MartinCamalich:2020dfe} and obtain the penguin amplitude at the electroweak scale,
\bea
\mathcal{L}_{bsa} (\mu_{EW}) & = & g_{bs}( \mu_{EW})\bar{s}P_R b \ a\,, \ \ \ {\rm with}~\nl
 g_{bs}(\mu_{EW}) & = & i\frac{\eta_t(\Lambda) m_b}{\ft} \frac{\sqrt{2}G_F m_t^2 V^*_{ts}V_{tb}}{16 \pi^2} \ln \frac{\Lambda^2}{m_t^2}\,,\
 \label{gbs}
\eea
where $\Lambda= 4 \pi \ft$ is the scale of new physics. We ignore the running of the WC to the scale $\mu \sim m_b$ where we do our phenomenology. This is justified in our analysis as the renormalization-group corrections are suppressed by $\alpha$ \cite{Bauer:2020jbp} and so $g_{bs}(\mu_{EW}) \approx g_{bs}( m_b)$. Including the loop term, the onshell ($b\to s a$) and offshell contributions ($b \to s a \to b \to s\bar{q}q$), where $q=u,d$, are given by
\bea
\mathcal{L}^{\rm onshell} & = & g_{bs}(\mu= m_b)\left[\bar{s}P_R b\right] a= \left[J_{b \rightarrow s}\right] a\,, \nl
\mathcal{L}^{\rm offshell}&=& \left[ J_{b \rightarrow s} \right]
\frac{\left[ \xi_d \frac{m_d}{\ft}  \bar{d} \gamma_5 d + \eta_u \frac{m_u}{\ft} \bar{u} \gamma_5 u\right]}
{m_{\pi^0}^2-m_a^2 + i m_a \Gamma_a}\,. \
\eea

\textbf{\emph{NP \bm{$\btopik$} fit}.--} To calculate the axion contributions to the $\btopik$ decays we can calculate an onshell and an offshell contribution. In the onshell case we assume there is mixing between the $a$ and the $\pi^0$ and so we can define a transformation between the gauge and the mass states as
\bea
\ket{a} & = & \cos \theta \ket{a_{\rm phy}} +  \sin \theta \ket{\pi^0_{\rm phy}}\,, \nonumber\\
\ket{\pi^0} & = &- \sin \theta \ket{a_{\rm phy}}+  \cos \theta \ket{\pi^0_{\rm phy}}\,. \
\eea
The onshell and offshell contributions to the ALP amplitude for $B  \to K a \to K \pi^0 $ give
\bea
\mathcal{A}&= &\mathcal{A}^{\rm onshell} + \mathcal{A}^{\rm offshell}\,, \ \ \ {\rm where}\\
\mathcal{A}^{\rm onshell} & = &   \bra{K}J_{b \rightarrow s}\ket{B} \langle{a}\ket{\pi^0_{\rm phy}}
 =  \bra{K}J_{b \rightarrow s}\ket{B} \sin {\theta}\,, \nl
\mathcal{A}^{\rm offshell} & = &  \frac{ m_{\pi^0}^2 \bra{K}J_{b \rightarrow s}\ket{B} }{m_{\pi^0}^2- m_a^2 + i m_a \Gamma_a}
\left[ \frac{ \eta_u f_\pi}{ 2 \sqrt{2}\ft}  - \frac{\xi_d f_\pi}{ 2 \sqrt{2} \ft}\right]\,. \nonumber
\label{ALPamp}
\eea
Here,
\bea
\bra{K}J_{b \rightarrow s}\ket{B} & = & g_{bs} \bra{K}\bar{s}P_R b\ket{B}\,, \\
\bra{K}\bar{s}P_R b\ket{B} &=& f_{+}(m_K^2)\frac{m_B^2-m_K^2}{2(m_b-m_s)} \nl
&& +~ f_{-}(m_K^2)\frac{m_K^2}{2(m_b-m_s)}\,,\\
\bra{K}\bar{s} \gamma^\mu b\ket{B} & = &  f_{+}(q^2)( p_B^\mu+ p_K^\mu) \nl
&& +~ f_{-}(q^2) ( p_B^\mu-p_K^\mu)\,.
\eea
We use the naive-factorization relations to calculate the offshell effect:
\bea
\bra{\pi^0} \bar{d} \gamma_5 d \ket{0}  &= & - \frac{f_\pi m_{\pi^0}^2}{ 2 \sqrt{2}  m_d}\,, \nonumber\\
\bra{\pi^0} \bar{u} \gamma_5 u \ket{0}  &=  & \frac{f_\pi m_{\pi^0}^2}{ 2 \sqrt{2}  m_u}\,, \nonumber\\
\ket{\pi^0} & = & \frac{\ket{ \bar{d} d}- \ket{\bar{u}u}}{ \sqrt{2}}\,.\
\label{MEfac}
\eea
Note that the effect of the offshell contribution can be absorbed in an effective mixing angle,
\bea
\sin {\theta} \to  \sin \theta + \frac{ m_{\pi^0}^2}{m_{\pi^0}^2- m_a^2}  \frac{\eta_u -\xi_d}{ 2 \sqrt{2}} {f_\pi \over \ft}\,.
\label{mix}
\eea
where we have assumed $\Gamma_a \ll |m_{\pi^0}- m_a|$. The term proportional to $m_{\pi^0}^2$  is the same as the $a-\pi^0$ mixing
 term usually discussed in the chiral Lagrangian description of the interaction of the ALP with mesons, with the ALP-quark couplings induced entirely by the ALP coupling to gluons; see for example Refs.~\cite{Bj_rkeroth_2018, Bauer:2021wjo}. A  similar mixing term proportional to $m_a^2$ is included in the onshell
 contribution to $\sin \theta$, i.e, in the first term on the right-hand-side of Eq.~(\ref{mix}).
If $\ft=1$~TeV
and $\sin\theta \sim 0.1$ with the mixing arising primarily from the second term, then $|\eta_u - \xi_d| \simeq 0.01$ gives $|m_{\pi^0}- m_a| \sim 1$~keV. Detecting an ALP so close in mass to the $\pi^0$ will pose a challenge for $B$ factories which have a $\pi^0$ mass resolution of a few MeV~\cite{Adachi:2018qme}.

With the ALP NP contribution added, we have for the $\btopik$ decays,
\bea
A^{+0}&=&-P_{tc}^\prime-P_{uc}^\prime e^{i\gamma}-\frac{1}{3}P_{EW}^{\prime C}\label{eq:ap0}\,,\\
\sqrt{2}A^{0+}&=&-T^\prime e^{i\gamma}-C^\prime e^{i\gamma}+P_{tc}^\prime - P_{uc}^\prime e^{i\gamma} \nl &&-~P_{EW}^\prime-\frac{2}{3}P_{EW}^{\prime C}+\mathcal{A}\label{eq:a0p}\,,\\
A^{-+}&=&-T^\prime e^{i\gamma}+P_{tc}^\prime-P_{uc}^\prime e^{i\gamma}-\frac{2}{3}P_{EW}^{\prime C}\label{eq:amp}\,,\\
\sqrt{2}A^{00}&=&-C^\prime e^{i\gamma}-P_{tc}^\prime+P_{uc}^\prime e^{i\gamma}-P_{EW}^\prime \nl
&&-~\frac{1}{3}P_{EW}^{\prime C}+\mathcal{A}\,,\label{eq:a00}
\eea
which satisfy Eq.~(\ref{eq:isorel}).
We simplify our fit by setting $P_{uc}^\prime = 0$ and taking the QCD-factorization inspired value for the ratio $|C^\prime|/|T^\prime| = 0.2$ \cite{Beneke:2009ek}. Typically to solve the puzzle we need $|\mathcal{A}|\sim\pewp\sim T'$. Notice that $T'$ is not
the dominant amplitude as it is suppressed by CKM elements.

A fit of just the SM amplitudes using the 4 branching ratios and the $A_{CP}(-+)$ measurement (which are the most constraining measurements) yields a $\chi^2/{\rm dof}=2.66/1$, and we are left with very large errors in the other $A_{CP}$ measurements. In fact, fitting to all observables other than the $A_{CP}(+0)$, in the SM we obtain $\chi^2/{\rm dof} =11.0/4 $ which is a poor fit. This requires us to include the ALP amplitude $\mathcal{A}$. The minimal fit that can be done to extract this amplitude is given in Table~\ref{tab:usedfit}. Any additional constraints on the system yield central values that differ by just a few percent.
\begin{table}[t]
\begin{center}
\begin{tabular}{|c|c|}\hline
Parameter & $|C^\prime|/|T^\prime|=0.2$\\
\hline
$\chi^2/{\rm dof}$&$3.64/3$\\ \hline
p-value & 30\% \\ \hline
$|T^\prime|$&$6.4\pm1.5$\\
$|P_{tc}^\prime|$&$50.30\pm0.47$\\
$|\mathcal{A}|$&$6.4\pm3.4$\\
$\delta_{C^\prime}$&$186\pm54$\\
$\delta_{P{tc}^\prime}$&$-18.1\pm5$\\
\hline
\end{tabular}
\caption{A fit of the SM amplitudes $T^\prime$ and $P_{tc}^\prime$, the relative phase of $C^\prime$, and the NP amplitude $\mathcal{A}$ with a fixed phase of $\pi/2$. The 8 measurements fit are the 4 branching ratios, $A_{CP}(-+)$, $A_{CP}(0+)$, $A_{CP}(00)$, and $S_{CP}(00)$; we do not fit $A_{CP}(+0)$ since it is independent of all parameters in the table. Magnitudes of the diagrammatic amplitudes are in eV and phases are in degrees. Note that the magnitudes and phases of the electroweak penguin diagrams are obtained using Eq.~(\ref{EWPrels}).}
\label{tab:usedfit}
\end{center}
\end{table}
This value of $|\mathcal{A}|$ allows us to evaluate $\sin\theta$ using Eq.~(\ref{ALPamp}). Using values of the masses taken from \cite{pdg}, the form factors from \cite{Khodjamirian:2017fxg}, and taking
$\ft=1$~TeV, we find
\begin{equation}
|\mathcal{A}|=i\eta_t(\Lambda)\sin\theta\left[5.71\times10^{-6}\ {\rm{GeV}}\right]\,.
\end{equation}
Using the value for $|\mathcal{A}|$ obtained from the fit, we have
\begin{equation}
\eta_t(\Lambda)\sin\theta = (1.12\pm0.60)\times10^{-3}\,.
\end{equation}

We now extract just the $\sin\theta$ term. The coupling $\eta_t(\Lambda)$ is unknown and cannot be properly extracted using this method. Note that if we just consider the branching ratio $B\rightarrow Ka$ we have
\begin{equation}
{\rm BR}(B\rightarrow Ka) = \frac{p_K\tau_B}{8\pi m_B^2}\frac{|\mathcal{A}|^2}{\sin^2\theta}\,.
\end{equation}

Using $|\mathcal{A}|$ obtained from the fit and branching ratio values $10^{-5}$ and $2\times10^{-5}$, we find
\bea
\sin\theta&=&0.188\pm0.029\,,\\
\sin\theta&=&0.133\pm0.021\,,
\eea
respectively. Note that a careful search of the decays $B \to K \pi^0$ around the $\pi^0$ mass may be able to observe the ALP as a diphoton resonance. We determine the value of $\eta_t(\Lambda)$ by using the value of $\sin\theta=0.133\pm0.021$:
\begin{equation}
\eta_t(\Lambda)=(8.4\pm4.7)\times10^{-3}\,.
\end{equation}

\textbf{\emph{\bm{$K\to\pi a$} amplitude}.--} We first consider the amplitude of $K^+\rightarrow \pi^+a$. This can come from $\pi^0-a$ mixing, so that $ K^+ \to \pi^+ \pi^0 \to \pi^+ a$. There is also direct production through the weak current~\cite{Bauer:2021wjo} which we can make small by an appropriate choice of ALP couplings to the light quarks $\xi_{d,s}$, and $\eta_u$.
Hence ${\rm BR}[K^+ \to \pi^+ a] \sim \sin^2\theta {\rm BR}[K^+ \to \pi^+ \pi^0]$ and this decay will be swamped by the $K^+ \to \pi^+ \pi^0$ decay. This is also the FCNC $s\rightarrow d$ transition that arises from a penguin loop (see for example Ref.~\cite{Izaguirre:2016dfi}) that contributes to $K^+\to \pi^+ a, K^0_L\rightarrow\pi^0a$ and $ K^0_S\rightarrow\pi^0a$. Using $\eta_t(\Lambda) = (8.4\pm4.7) \times10^{-3}$, we obtain the branching ratios in the penguin generated $K\rightarrow\pi a$ decays:
\begin{align}
{\rm BR}(K^+\rightarrow\pi^+a)&=(4.2\pm3.3)\times10^{-8}\,,\\
{\rm BR}(K^0_L\rightarrow\pi^0a)&=(1.8\pm1.4)\times10^{-7}\,,\\
{\rm BR}(K^0_S\rightarrow\pi^0a)&=(5.5\pm4.3)\times10^{-11}\,.
\end{align}
Other constraints from the $B$ and the $K$ system are discussed in Ref.~\cite{Datta:2019bzu} in which a  model with similar structure and parameters has been considered.

\textbf{\emph{\bm{$D$} system}.--} We now consider the contribution of the ALP to the $D\to K\pi$ system. In our model the ALP
enters a meson decay process only via a one-loop penguin diagram when the final state has a $\pi^0$. Based on the CKM
matrix elements that enter in the amplitude, $D$-meson decays can be broadly categorized into Cabibbo-favored
[$\propto V^*_{cs}V_{ud}$], single-Cabibbo-suppressed (SCS) [$\propto V^*_{cd}V_{ud}$ or $\propto V^*_{cs}V_{us}$], and doubly-Cabibbo-suppressed [$\propto V^*_{cd}V_{us}$]. Of these, a penguin diagram can only appear in the SCS
$D$ decays, and only three of these involve a $\pi^0$ in the final state ($D^0\to\pi^0\pi^0, D^+\to\pi^+\pi^0, D^+_s\to
K^+\pi^0$). The ALP-penguin amplitude $\mathcal{A}$, introduced earlier in $B$-decays, also contributes to each of these decays.
A key difference is that in $D$ decays the bottom quark, rather than the top quark, runs in the penguin loop. We denote
this new amplitude by $\mathcal{A}^D$.

Since the penguin diagram here is similar to the diagram that contributes to the $B$ decays, we obtain similar
expressions for the ALP contribution in the $D$ system. Key changes appear in the quark flavors, since instead
of a $b\to s$ transition, now we have a $c\to u$ transition:
\bea
&&\mathcal{A}^D~=~g_{cu}\Braket{P|J_{c\rightarrow u}|D}\Braket{a|\pi^0_{\rm phy}}\,,\\
&&g_{cu}~=~\frac{\xi_{b}(\Lambda)m_c}{\ft}\frac{\sqrt{2}G_Fm_b^2V_{cb}^*V_{ub}}{16\pi^2}\ln\frac{\Lambda^2}{m_b^2} \nl
&&~~~~~=~\xi_b(\Lambda)e^{-i\gamma}\left[5.81\times10^{-12}\,{\rm {GeV}}\right]\,,\\
&&\Braket{P|J_{c\rightarrow u}|D}~=~f_+^{D\rightarrow P}(m_P^2)\frac{m_D^2-m_P^2}{2(m_c-m_{u})} \nl
&&~~~~~~~~~~~~~~~~~~~~-~f_-^{D\rightarrow P}(m_P^2)\frac{m_P^2}{2(m_c-m_{u})}\,,
\eea
where
\begin{align}
\Braket{\pi|J_{c\rightarrow u}|D}&=6.52 ~{\rm GeV}\,,\\
\Braket{K|J_{c\rightarrow u}|D}&=6.43 ~{\rm GeV}\,,\\
\Braket{K|J_{c\rightarrow u}|D_s}&=6.26 ~{\rm GeV}\,.
\end{align}
The phase $\gamma$ arises from the CKM matrix element $V_{ub}$. Now, with the value of $\sin\theta$ obtained from the $B$ decays,
assuming $\xi_b(\Lambda) < \eta_t(\Lambda)$, and $\ft=1$ TeV, we find that the ALP contribution to the $D$ decays,
\begin{equation}
|\mathcal{A}^D| < 5 \times 10^{-14} ~{\rm GeV}\,.
\end{equation}
This amplitude is several orders of magnitude smaller than the typical SM contribution in SCS decays, which are of the order
of $10^{-7}$ GeV \cite{Bhattacharya:2012ah}. We, therefore, conclude that the ALP contribution does not significantly affect
$D\to K\pi$ branching ratios. In Table \ref{tab:SCSD} we provide the experimental values for the magnitudes of the decay
amplitudes (calculated from the measured branching ratios) and the direct-CP asymmetries.
\begin{table}[t]
\centering
\begin{tabular}{|c|c|c|c|}
\hline
Process & Expt.~$|A|$ ($\times10^{-7}$ GeV) & Expt.~$A_{CP}$ (\%) \cite{pdg} \\ \hline
$D^0\to\pi^0\pi^0$ & $3.54\pm0.05$ \cite{pdg} & $0\pm0.6$ \\ \hline
$D^+\to\pi^+\pi^0$ & $2.738\pm0.006$ \cite{pdg} & $2.4\pm1.2$ \\ \hline
$D^+_s\to K^+\pi^0$& $1.9\pm1.1$ \cite{CLEO:2009fiz} & $-26.6\pm23.8$ \\ \hline
\end{tabular}
\caption{The magnitudes of measured amplitudes and direct-CP asymmetries in SCS $D$-meson decays. Only included are processes
in which the ALP contributes.}
\label{tab:SCSD}
\end{table}
We now estimate the contribution of the ALP to the direct-CP asymmetries in SCS $D$-decays as follows. The generic
$D$-decay amplitude in the presence of the ALP can be expressed as,
\beq
\label{eq:ALPD}
A_{D\to K\pi} ~=~ |a_{\rm SM}| e^{i\delta}e^{i\phi} + i|\mathcal{A}^D|\,,
\eeq
where $a_{\rm SM}$ is the magnitude of the SM part of the decay amplitude, $\delta$ is the relative strong phase, and $\phi$ is
the relative weak phase between the SM part and the ALP contribution. This leads to the direct-CP asymmetry,
\beq
A_{CP} ~=~ \frac{2 x \sin\delta\cos\phi}{1 + x^2 + 2 x \cos\delta\sin\phi}\,,
\eeq
where $x = |\mathcal{A}^D|/|a_{SM}| \lesssim 10^{-7}$. Clearly, a nonzero CP asymmetry can appear even if the SM term has a small weak phase.
This property is due to the $i$ in the coefficient of the ALP term in Eq.~(\ref{eq:ALPD}) which changes sign under CP conjugation.
Also, since $x \lesssim 10^{-7}$, the ALP's contribution to $A_{CP}$ in $D$-decays is several orders of
magnitude below the current sensitivity of flavor experiments.

\textbf{\emph{\bm{$(g-2)_{\mu,e}$} anomalies}.--} Our scenario can be easily extended to explain the anomalies in the anomalous magnetic moments $a_\ell = (g-2)_\ell/2$ of the muon and electron. We consider the $4.2\sigma$ $a_\mu$ anomaly from a combination of the BNL and Muon g-2 experiments \cite{g-2mu} with
\begin{eqnarray}
\Delta a_\mu &=& a^{\rm exp}_\mu-a^{\rm SM}_\mu = (251 \pm 59)\times 10^{-11}\,.
\end{eqnarray}
There are two values of $a_e$ which are inferred from measurements of the fine structure constant, and that are inconsistent with each other.
The $a_e$ value obtained from
Laboratoire Kastler Brossel~\cite{Morel:2020dww} and Berkeley~\cite{Parker:2018vye}
measurements of the fine-structure constant yield~\cite{Aoyama:2012wj,Aoyama:2019ryr,Hanneke:2008tm}
\begin{eqnarray}
\Delta a^{\rm LKB}_e &=& a^{\rm exp}_e-a^{\rm LKB}_e = (4.8\pm 3.0) \times 10^{-13}\,, \nonumber \\
\Delta a^{\rm B}_e &=& a^{\rm exp}_e-a^{\rm B}_e = (-8.8\pm 3.6) \times 10^{-13}\,.
\end{eqnarray}

Under our assumption that $\kappa \ll \sin\theta\,g_{\pi\gamma\gamma}$, the couplings of the ALP to the muon and electron must be $\xi_\mu m_\mu/\ft \sim 10^{-5}$ and $\xi_e m_e/\ft \sim 10^{-7}$ for $\Delta a^{\rm LKB}_e$~\cite{Keung:2021rps}
so the loop-induced  $Za\gamma$ coupling remains small. The values of $\xi_\mu m_\mu/\ft \sim 10^{-5}$ and $\xi_e m_e/\ft \sim -10^{-6}$ for $\Delta a^{\rm B}_e$
give a too large $Za\gamma$ coupling.

\textbf{\emph{Summary}.--} In perhaps a first analysis with ALPs in hadronic $B$ decays, we have proposed a new solution to the $\btopik$ puzzle with an ALP with mass close to the $\pi^0$ mass. Our solution preserves the isospin relation in Eq.~(\ref{eq:isorel}), and is consistent with constraints from $B$, $K$, and $D$ decays.
We point out that this ALP can also explain the $g-2$ anomalies of the muon and electron. A careful scan of the decay products in $B \to K \pi^0$ around the $\pi^0$ mass may reveal the ALP.

\bigskip
\noindent
\textbf{\emph{Acknowledgments}.--}
This work was supported by NSF Grant No. PHY-2013984 (B.B.), PHY1915142 (A.D.), DOE Grant No. de-sc0010504 (D.M.), and SERB,
Govt. of India, under the grant CRG/2018/001260 (S.N.). B.B. thanks D.~London for useful conversations.
\bibliography{ref_Kpiaxion}

\end{document}